\newcommand*{\ATLASLATEXPATH}{./}

\documentclass[cernpreprint,texlive=2011,onecolumn,USenglish]{\ATLASLATEXPATH atlasdoc}

\pdfoutput=1

\usepackage [compress,comma,numbers]{natbib}

\usepackage[biblatex=false]{\ATLASLATEXPATH atlaspackage}

\usepackage{\ATLASLATEXPATH atlascontribute}

\usepackage{\ATLASLATEXPATH atlasphysics,\ATLASLATEXPATH atlasheavyion, \ATLASLATEXPATH atlasprocess}

\graphicspath{{./}}

\usepackage{pPb_zll_paper-defs}

\hypersetup{pdftitle={ATLAS draft},pdfauthor={The ATLAS Collaboration}}

\AtlasTitle{$Z$ boson production in \pPb collisions at \energy\ measured with the ATLAS detector}

\author{The ATLAS Collaboration}

\AtlasRefCode{HION-2013-09}
\PreprintIdNumber{CERN-PH-2015-146}
\AtlasJournal{Physical\ Review\ C}
\AtlasAbstract{
The ATLAS Collaboration has measured the inclusive production of $Z$ bosons via their decays into electron and muon pairs in \pPb\ collisions at \energy\ at the Large Hadron Collider. The measurements are made using data corresponding to integrated luminosities of 29.4 \inb\ and 28.1 \inb\ for \Zee and \Zmm, respectively.  The results from the two channels are consistent and combined to obtain a cross section times the \Zll\ branching ratio, integrated over the rapidity region $|\ystarZ|<3.5$, of 139.8 $\pm$ 4.8 (stat.) $\pm$ 6.2 (syst.) $\pm$ 3.8 (lumi.) nb.  Differential cross sections are presented as functions of the $Z$ boson rapidity and transverse momentum, and compared with models based on parton distributions both with and without nuclear corrections.  The centrality dependence of $Z$ boson production in \pPb\ collisions is measured and analyzed within the framework of a standard Glauber model and the model's extension for fluctuations of the underlying nucleon-nucleon scattering cross section.}

\begin{document}

\maketitle

\tableofcontents

\section{Introduction}
The study of electroweak bosons in \PbPb\ collisions at the Large Hadron Collider (LHC) at CERN has demonstrated that 
 the production rate of non-strongly interacting particles scales with the number of nucleon-nucleon collisions, {$N_{\mathrm{coll}}$.  This has been observed for photons \cite{cms_photons}, $W$ bosons \cite{cms_ws,wboson_paper}, and $Z$ bosons \cite{PhysRevLett.110.022301,cms_z}.  The momentum and rapidity distributions of $Z$ boson yields  are consistent with {\scshape Pythia}  \cite{Sjostrand:2007gs} simulations of \pp\ collisions 
multiplied by the average nuclear thickness function, \mTAA , which is equivalent to \avgNcoll divided by the total nucleon-nucleon cross section \cite{PhysRevLett.110.022301}.  
$Z$ boson production in \PbPb\ collisions was found to be consistent with Next to Leading Order perturbative Quantum Chromodynamics (NLO QCD) calculations that disregard nuclear modifications in the treatment of parton distribution functions (PDF).  However, nuclear modification is not excluded within the precision of the measurement  \cite{hannuDIS}.  The production of $Z$ bosons, when  examined as a function of centrality, 
was also found to scale with \avgNcoll.

To differentiate between initial- and final-state effects in heavy ion (HI) collisions, the study of \pPb\ collisions is used at the LHC. One could expect that the hot and dense QCD medium cannot be formed in such collisions, unlike in the \PbPb\ case, and that modifications to the final-state particles relative to nucleon-nucleon collisions  should originate from the initial state of the nucleus. This assumption was challenged by the very first results from \pPb\ collisions at \energy\ produced at the LHC in 2012. Results on multi-particle correlations, published by three LHC experiments~ \cite{alice_pPb_flow1,cms_pPb_flow1,atlas_pPb_flow1,alice_pPb_flow2,cms_pPb_flow2,atlas_pPb_flow2,atlas_pPb_ridge}, revealed collective behavior in \pPb\ collisions similar to that previously measured in HI collision systems. 
The yields of jets measured by ATLAS  scale with \avgNcoll when measured inclusively for all centralities, but show significant deviations from binary scaling when considered in centrality selections \cite{ATLAS:2014cpa}.  The CMS Collaboration has measured dijet pseudorapidity distributions and found them to agree better with predictions that include nuclear PDF modifications than with predictions that do not include nuclear effects  \cite{CMS_pPbdijet_paper}. 
The CMS Collaboration has also recently measured production of $W$ bosons in \pPb\ collisions and observed hints of nuclear modifications  of the PDF \cite{CMS_W_pPb}.   Collectively, these results have highlighted the need for a better understanding of the initial conditions of \pPb\ collisions.

Unlike symmetric \PbPb\ collisions, in \pPb\ collisions nuclear modifications of the PDFs in the lead nucleus create an asymmetry in the rapidity-dependent cross section of $Z$ bosons; this presents an attractive observable for the study of initial-state nuclear conditions.  The centrality-dependent yield of $Z$ bosons is a well suited probe to test our understanding of \pPb\ collision geometry.     
The LHCb Collaboration has made a first exploratory measurement of $Z$ bosons at far forward and backward rapidities \cite{lhcb_Z_pPb} based on an integrated luminosity of 1.6 \inb.  

This paper presents the results of the measurement of $Z$ boson production in \pPb\ collisions at \energy\ using the ATLAS detector. The yield of $Z$ bosons is measured as a function of their transverse momentum \ptZ, rapidity in the center-of-mass frame (\ystarZ),  
\footnote{ATLAS uses a right-handed coordinate system with its origin at the nominal interaction point (IP) in the center of the detector and the $z$-axis along the beam pipe. The $x$-axis points from the IP to the center of the LHC ring, and the $y$-axis points upward. Cylindrical coordinates $(r,\phi)$ are used in the transverse plane, $\phi$ being the azimuthal angle around the beam pipe. The rapidity in the laboratory frame is given by $y^{\mathrm{lab}} = \frac{1}{2}\ln\frac{E+p_z}{E-p_z}$ and pseudorapidity is defined as $\eta = - \ln[\tan(\theta/2)]$.  Positive rapidity corresponds to the direction the proton beam travels, ``proton-going", and negative rapidity is referred to as ``Pb-going".  In this convention the asymmetric beam energies result in a center-of-mass shifted to rapidity $y^{*}=y^{\mathrm{lab}}-0.465$.}
 and centrality.
The leptonic decays of the $Z$ boson (\Zee\ and \Zmm) are used for its reconstruction. In the muon channel it is possible to reconstruct the $Z$ boson in the rapidity range $-3<\ystarZ<2$, while in the electron channel this range can be extended to $|\ystarZ|<3.5$.
The larger acceptance in rapidity for \Zee\ candidates is possible because of the larger acceptance of the ATLAS calorimeters compared to the muon spectrometer (MS; Secs.~\ref{sec:detector} and \ref{sec:lep_reco}).  The efficiency of $Z$ boson reconstruction is calculated from detector simulations (Sec.~\ref{sec:MC}).  Backgrounds in each channel are estimated using simulations and data-driven methods (Sec.~\ref{sec:yield_extract}).  Results measured in the dimuon and dielectron decay channels are combined after accounting for uncertainties and their correlations (Sec.~\ref{sec:combination}).  
The measured cross sections and centrality-dependent yields are compared to models of \pPb\ collisions composed of 82 \pp\ and 126 \pn\ collisions in which the production of $Z$ bosons is obtained using perturbative QCD calculations. 

\section{The ATLAS detector \label{sec:detector}}
The ATLAS detector~\cite{Aad:2008zzm} at the LHC covers nearly the entire solid angle around the collision point. It consists of an inner tracking detector surrounded by a thin superconducting solenoid, electromagnetic and hadronic calorimeters, and the MS.
	
The inner-detector (ID) system is immersed in a 2~T axial magnetic field and provides charged particle tracking in the pseudorapidity range $|\eta|<2.5$. It comprises a high-granularity silicon pixel detector covering the collision region, surrounded by a silicon microstrip tracker and a transition radiation tracker.  

The calorimeters cover the range $|\eta|< 4.9$. Within the region $|\eta|< 3.2$, electromagnetic calorimetry is provided by barrel and endcap high-granularity lead liquid-argon (LAr) calorimeters, with an additional thin LAr presampler covering $|\eta| < 1.8$.  Behind the electromagnetic calorimeter there is a steel/scintillator  sampling  hadronic  calorimeter
covering $|\eta| < 1.7$, and LAr hadronic calorimeters extend the coverage to $|\eta| < 4.9$.  Forward electromagnetic calorimeters (FCals) are located in the range  $3.1<|\eta|<4.9$.  Electrons may be reconstructed over the entire electromagnetic calorimeter system, $|\eta|< 4.9$.

The MS comprises separate trigger and high-precision tracking chambers that measure the deflection of muons in a magnetic field generated by superconducting air-core toroids. The precision chambers cover the region $|\eta| < 2.7$ with three layers of monitored drift tubes~(MDT) chambers, complemented by cathode strip chambers~(CSC) in the innermost layer of the forward region. The muon trigger system covers the range $|\eta| < 2.4$ with resistive plate chambers~(RPC) in the barrel ($|\eta|< 1.05$), and thin gap chambers~(TGC) in the endcap regions ($1.05<|\eta|< 2.4$).

The ATLAS detector has a three-level trigger system~\cite{Aad:2012xs}: the hardware-based level-1 (L1) trigger and the software-based High Level Trigger (HLT), which is subdivided into the Level-2 (L2) trigger and Event Filter (EF). Single-electron and single-muon triggers 
are used to acquire the data analyzed in this paper.  Minimum-bias events are selected based on signals in the minimum-bias trigger scintillators (MBTS) that detect charged particles in the range $2.1<|\eta|< 3.9$. 

\section{Analysis}
\label{sec:ana}
\subsection{Data sample}
\label{sec:selection}
This analysis uses the 2013 ATLAS \pPb\ collision data at \energy, produced from a 4 TeV proton beam and a 1.57 TeV per nucleon lead beam.  The asymmetric energy of the beams resulted in a shift of the center-of-mass by 0.465 units of rapidity relative to the laboratory frame.   After 60\% of the data were recorded the directions of the proton and lead beams were reversed.  Results obtained in the two data periods are found to be consistent with each other.  In this paper all data from both periods are presented using the convention that the proton beam travels \textit{forward} in the positive rapidity direction.   

Following stringent data-quality requirements, the \Zee\ and \Zmm\ analyses use samples corresponding to integrated luminosity values of 29.4 $\pm$ 0.8 \inb\ and 28.1 $\pm$ 0.8 \inb, respectively.  
The luminosity measurement for the
2013 \pPb\ data is calibrated based on dedicated beam-separation scans. Systematic uncertainties similar to those studied for the calculation of \pp\ luminosity \cite{lumi_paper} are calculated.  The combination of these systematic uncertainties results in a total uncertainty in the ATLAS luminosity scale during proton-lead collisions at \energy\ of 2.7\%.  Minimum-bias \pPb\ collisions were selected by a trigger based on a signal in both MBTS counters.  Minimum-bias events are required to have the time measured in each  MBTS be consistent within 10 ns, and a reconstructed collision vertex within 175 mm of the nominal collision point in the longitudinal direction \cite{pPb_multpaper}.  

\subsection{Lepton reconstruction  \label{sec:lep_reco}}
Electron candidates are first identified by the L1 trigger as a cluster of cells in the electromagnetic calorimeter, formed into $(\Delta\phi\times\Delta\eta)=0.1\times0.1$ trigger towers, within the range $|\eta|<2.5$ and with the cluster transverse energy exceeding 5 GeV.  The HLT then incorporates tracking information from the ID and imposes electron identification requirements on the electron candidates.  A trigger for at least one electron candidate with $\ET>15$ \GeV\ and satisfying loose identification requirements is used to select events.   

In the offline analysis, electron candidates within $|\eta|<4.9$ are selected using the ATLAS reconstruction algorithm~\cite{eperf_2014paper}.  Electrons with $|\eta|<2.47$, referred to as midrapidity electrons, require the matching of a track to an energy cluster in the electromagnetic calorimeter.    In addition to the reconstruction requirements, further electron identification selections based primarily the shower shape in the electromagnetic calorimeter are made to reject background.   Electron identification requirements used in previous ATLAS analyses ~\cite{eperf_2014paper} are used to provide quality classification of electrons based on the tightness of the identification criteria they satisfy.  The triggering electron is required to have $\et>20$\,\GeV, to be outside the pseudorapidity interval $1.37<|\eta|<1.52$ (a transition region between the barrel and endcap calorimeters which contains a relatively large amount of inactive material), and to satisfy tight identification quality requirements.  If the other electron is within $|\eta|<2.47$, it must have $\et>10$ \GeV\ and satisfy looser quality requirements.  Forward-electrons are those reconstructed within the range $2.5<|\eta|<4.9$ based on energy deposited in the FCal \cite{eperf_2014paper}.  There is no tracking in this region, so the electron candidate reconstruction and identification is derived solely from the calorimeter signal and does not have an associated charge.  Forward-electrons are required to have $\et>20$ \GeV.  Of \Zee\ decays with $|\ystarZ|<3.5$, approximately 82\% fall into the fiducial acceptance defined by the electron \pt\ and $\eta$ requirements.    

Muon candidates are first identified at the L1 trigger, based on hits in either the RPC or TGC.  The HLT then reconstructs muon tracks in the vicinity of the detector region reported by the L1 trigger. The L2 trigger uses an algorithm to perform a fast reconstruction of muons, which is then refined in the EF by incorporating the hits in the ID tracking as well as those in the MS tracking.  
Events containing at least one muon with \pt\ greater than 8 GeV are accepted by the HLT.

For the \Zmm\ analysis, muons are identified from candidates reconstructed in both the MS and ID \cite{muon_perf}.  Muons are reconstructed separately in the MS and ID and a $\chi^{2}$-minimization procedure is used to obtain combined muon kinematic information.  To reduce background from jets, each muon is required to pass a loose track-based isolation selection.  Tracks are considered in a cone of size $\Delta R=\sqrt{(\Delta\eta)^{2}+(\Delta\phi)^{2} } = 0.3$ around the direction of the muon.  The muon is considered isolated if the scalar sum of the \pT\ of these tracks, excluding the muon, is less than 50\% of the muon \pT.  The efficiency of this selection is greater than 99\%.  The triggering muon is required to have $\pt>20$ \GeV\ and be within $|\eta|<2.4$.  The second muon must be within $|\eta|<2.47$ and have $\pt>10$ \GeV.  Approximately 68\% of $Z$ bosons with $-3<\ystarZ<2$ fall  into the fiducial acceptance defined by the muon kinematic requirements.

\subsection{Centrality}
\label{sec:centrality}
In addition to measuring the $Z$ boson cross section, the $Z$ boson yield per minimum-bias event is measured for different centrality selections.
In order to characterize the \pPb\ collision geometry, each event is assigned a centrality based on the total transverse energy measured in the FCal on the  Pb-going side ($-4.9<\eta<-3.2$), \sfcal\ \cite{pPb_multpaper}.  
Collisions with more (fewer) participating nucleons are referred to as central (peripheral).  
As in Ref.  \cite{pPb_multpaper}, the standard Glauber model \cite{Glauber} approach is used to calculate the mean number of participating nucleons, \avgNpart.  The mean number of inelastic nucleon-nucleon collisions, \avgNcoll, is $\avgNpart -1$.  Based on the observed centrality dependence of the charged particle multiplicity \cite{pPb_multpaper}, a Glauber-Gribov Color Fluctuation (GGCF) model  \cite{gribov1,gribov2}, an extension to the Glauber model which allows event-by-event fluctuations of the nucleon-nucleon cross section  $\sigma(N+N\rightarrow X)$, is also considered.  In this model the magnitude of the fluctuations is characterized by the parameter $\omega_{\sigma}$, with $\omega_{\sigma}$=0 corresponding to the standard Glauber model.  Following Ref.~\cite{pPb_multpaper}, two values of $\omega_{\sigma}$, 0.11 and 0.2, based on the calculations in Refs.~\cite{gribov1,gribov2}, are implemented and considered for this analysis.  

Following Ref.~\cite{pPb_multpaper}, the centrality selections used for this analysis, in order from most central to most peripheral, are 0--5\%, 5--10\%, 10--20\%, 20--30\%, 30--40\%, 40--60\%, and 60--90\%.  For the study of the \ystarZ\ distribution as a function of centrality (see Sec.~\ref{sec:cent_yields}), larger bins are used: 0--10\%, 10--40\%, and 40--90\%.  For the most peripheral collisions, centrality greater than 90\%, centrality modeling, and associated geometric quantities are not well constrained.  Pileup events, those containing multiple \pPb\ interactions from the same bunch-crossing, are removed by rejecting events in which more than one primary vertex is reconstructed.  The fraction of candidate events which is removed from the centrality-selected $Z$ boson yield analysis due to pileup rejection is approximately 5\%.  Diffractive events are identified by a rapidity gap (defined by the absence of calorimeter energy clusters) of greater than two units on the Pb-going side of the detector, and excluded.  This leads to a rejection of less than 0.1\% of $Z$ boson candidate events.   
The total number of minimum-bias events corresponding to the luminosity sampled by the trigger, $N_{\mathrm{evt}}$, is used to define the $Z$ boson yield per event in each centrality selection.

Besides its sensitivity to the event geometry, the \sfcal\ for a fixed geometry may also be affected by the presence of a hard scattering process in the event.
In particular, the calculation of centrality for \Zee\ events in which there is a forward-electron in the Pb-going side FCal, is biased by the energy of the electron.  This is corrected by subtracting the transverse energy of the electron from \sfcal.  The  subtraction procedure is found to effectively recover the correct centrality of minimum-bias events into which simulated  \Zee\ decays containing electrons in the Pb-going side FCal were overlaid.

In addition to the case where there is a $Z$-decay electron in the Pb-going side FCal, \sfcal\ may be more subtly biased in all $Z$ boson events.  The presence of a $Z$ boson (or any hard process) is correlated with a higher transverse energy of the underlying event. Consequently, more energy may be deposited in the Pb-going side FCal in events containing a hard scattering process than in those that do not contain one. This causes a bias in the centrality-dependent yield, as the $Z$ boson yield is enhanced in the more central events but depleted in the more peripheral ones.  This effect, referred to as a centrality bias,  has been noted for yields of hard processes in \dAu collisions at $\sqn=200$~GeV by the PHENIX Collaboration  \cite{PHENIX_dAu_cent}, and by the ALICE Collaboration in \pPb\ collisions at \energy\ by the ALICE Collaboration \cite{ALICE_pPb_cent}.

A correction to the centrality-dependent yields of hard processes in \dAu collisions at $\sqn=200$~GeV has been studied by the PHENIX Collaboration \cite{PHENIX_dAu_cent} (the correction is also calculated for \pPb\ at \energy).  The centrality bias correction used by the PHENIX Collaboration is based on the modeling of an increase in the mean particle multiplicity produced by the specific  \NN collision that undergoes a hard scattering.  Recently, similar calculations of a centrality bias have been made in which all \NN collisions may contribute to an increase in the particle multiplicity \cite{dennis_peter_paper}.   The increase in multiplicity stemming from each \NN collision is taken to be proportional to the contribution from that collision to the \ET in the event.  
This model is applied to the ATLAS \pPb\ centrality classification for the standard Glauber analysis as well as the GGCF models, and thus used to calculate corrections to the hard process yield measured in a given centrality bin.  Because the $N_{\text{part}}$ probability distribution varies less steeply in the GGCF models than in the standard Glauber model, the centrality bias corrections are closer to unity for the GGCF cases.  The corrections from Ref.~\cite{dennis_peter_paper} are shown in Figure \ref{fig:cent_corrections}.  
The reciprocals of the corrections are applied as multiplicative factors to the centrality-dependent $Z$ boson yields measured in the present analysis.

\begin{figure}[htb]
\begin{center}
\includegraphics[width=0.5\textwidth]{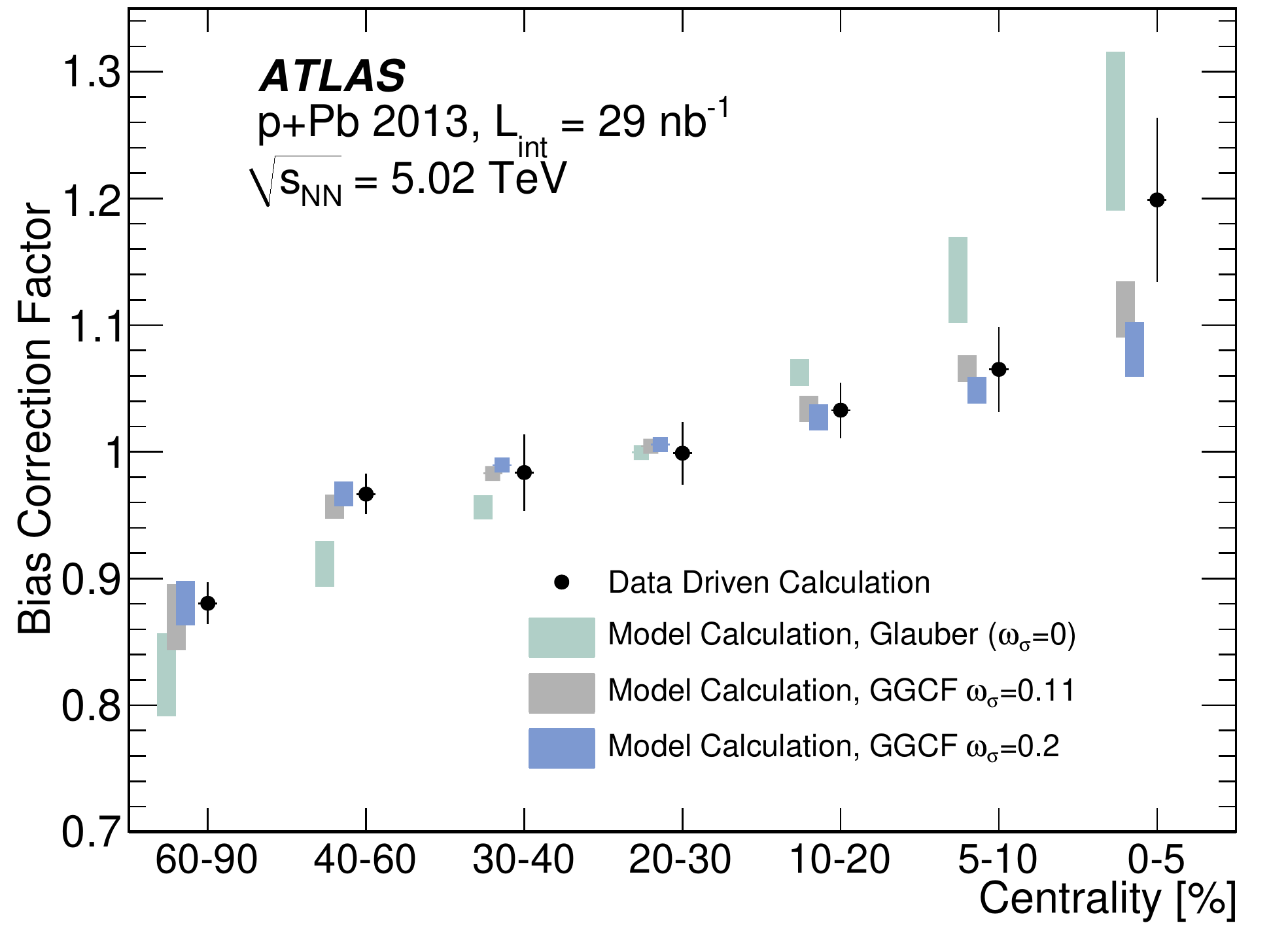}
\caption[Centrality bias corrections]{(Color online) Centrality bias correction factors and their uncertainties (bars), from Ref.~\cite{dennis_peter_paper}, for the Glauber and GGCF model configurations, and the bias correction factors derived from data (points) as explained in the text. The reciprocals of the correction factors are applied as multiplicative factors to the centrality-dependent Z boson yields.
}
\label{fig:cent_corrections}
\end{center}
\end{figure}

Using $Z$ bosons measured in ATLAS, data-driven centrality bias corrections may be calculated and compared with the results from Ref.~\cite{dennis_peter_paper}.  
To do so, the corrections are calculated by comparing the transverse energy deposited in the FCal in events selected by the minimum-bias trigger and in $Z$ boson events from \pp\ collisions (in which there is no centrality to consider).  This effect is studied in \pp\ collisions at $\sqrt{s}=2.76$~TeV and $\sqrt{s}=7$~TeV from the 2011 LHC run.  At both energies a significant increase in the mean transverse energy deposited in the FCal is observed, and within the uncertainties of the measurement found to be independent of the $Z$ boson kinematics.  A single value is interpolated from the $\sqrt{s}=2.76$~TeV and $\sqrt{s}=7$~TeV data for \energy.  The interpolation is performed using both logarithmic and linear expressions, and the difference between them contributes to the systematic uncertainty.  A correction is made to account for the shifted center-of-mass in the \pPb\ system which changes the effective FCal acceptance in $\eta$ compared to \pp.  From this procedure an additive shift to \sfcal\ of 2.0 $\pm$ 0.5 GeV is calculated.
In each \pPb\ \Zll\ event this value is subtracted from \sfcal\ and the resulting value used to determine a corrected centrality of the event.  The centrality-dependent yield may be constructed, according to the method described in Sec.~\ref{sec:yield_extract}, using both the subtracted and unsubtracted \sfcal\ values and their ratio is then comparable to a centrality bias correction factor that  may be compared to those calculated in Ref.~\cite{dennis_peter_paper}.  As shown in Figure \ref{fig:cent_corrections}, within the uncertainties this data-based method is compatible with the model calculations. 

\subsection{Monte Carlo simulation corrections\label{sec:MC}}
The trigger, reconstruction, and identification efficiencies of electrons and muons as well as the muon isolation efficiency are evaluated by a Monte Carlo  (MC) simulation complemented by data-driven estimates of these quantities.  Using the {\scshape Powheg} generator \cite{powhegz} (with the CT10 PDF \cite{CT10_paper}) interfaced to {\scshape Pythia}8 \cite{Sjostrand:2007gs} for simulation of the parton shower, approximately 10 million \Zee\ and 4 million \Zmm\ events were simulated.  The $Z$ bosons were generated from $pp$ and $pn$ collisions, which were added together with weights 82/208 and 126/208, respectively, corresponding to the numbers of protons and neutrons in the Pb ion. The response of the ATLAS detector to the generated particles was modeled using GEANT4~\cite{GEANT4_paper,ATLAS_sim}.  Due to the dependence of electron identification and reconstruction efficiency on detector occupancy, the simulated  \Zee\ events were overlaid with data events selected with the minimum-bias trigger, and then reconstructed.  

To cross-check the efficiencies calculated in the MC simulation, a `tag-and-probe' technique is employed.  A `tag' is defined as a fully reconstructed high-quality triggered lepton, whereas the `probe' is a lepton candidate to which triggering, reconstruction, or quality requirements are not applied.  Using tag-and-probe pairs with an invariant mass $m_{\ell\ell}$  consistent with selection of $Z$ bosons, the efficiency of the probe with additional requirements is calculated.  The mass window used depends on the background present in the probe sample, and ranges from $80<m_{\ell\ell}<100$ \GeV\ to $87<m_{\ell\ell}<95$ \GeV. For example, the electron trigger efficiency is measured from high-quality reconstructed electron probes selected without an \textit{a priori} trigger requirement, and the MS reconstruction efficiency is measured from charged-particle tracks in the ID without an \textit{a priori} MS signal requirement.  The MC simulation is scaled to match the efficiencies determined with the data-driven tag-and-probe method.
The factors used to scale the MC electron response are derived from the 2013 \pPb\ data set.  Muon reconstruction is insensitive to the differences between 2013 \pPb\ and 2012 \pp\ conditions and therefore the scale factors for muons are taken from \Zmm\ events collected in the 2012 \pp\ dataset \cite{muon_perf}.  The scale factors for the \Zee\ (\Zmm) MC events deviate from unity by less than 5\% (1\%).

The trigger efficiencies of electrons and muons depend on $\eta$, and have average values of approximately 95\% and 83\%, respectively.  The reconstruction and identification efficiency of the more stringently selected electrons is approximately 75\%, whereas the reconstruction and identification efficiency for the looser-quality midrapidity electrons is approximately 88\% and the forward-electron efficiency is about 65\%.  These values depend on $\eta$ and \pt.  The muon reconstruction efficiency is approximately 95\%, depending on $\eta$.  

Correction factors for the yields of \Zee\ and \Zmm\ candidates are calculated from the MC simulations as functions of \ptZ, \ystarZ, and \pPb\ centrality.  These corrections take into account the cumulative losses due to trigger, reconstruction, and identification efficiency as well as the kinematic acceptance of the decay leptons.  The correction is defined relative to all generated $Z$ bosons within the mass window $66<m_{\ell\ell}<116$ \GeV.  The total efficiency for reconstructing a produced $Z$ boson, including acceptance, is approximately 55\% for \Zee\ and 65\% for \Zmm.  Following the subtraction of background (see Sec~\ref{sec:yield_extract}) and application of the correction factor, a corrected yield of $Z$ bosons is obtained in each bin of \ptZ\ and \ystarZ.  The uncertainty on the correction factor follows from the uncertainty of the data-driven tag-and-probe checks of the MC, primarily due to the relatively low number of tag-and-probe events in the data.  The uncertainty associated with the lepton identification efficiency is the dominant uncertainty for both the \Zee\ and \Zmm\ analyses.  This uncertainty is approximately 10\% for \Zee\ (rising as high as 15\% for pairs including a forward-electron), and 1.5\% for \Zmm.   The sizable uncertainty for \Zee\ is primarily driven by the limited size of the tag-and-probe 2013 \pPb\ dataset. The other uncertainties are typically less than 2\% in both channels.

\subsection{Yield extraction  \label{sec:yield_extract}}
To form \Zee\ candidates, all electrons found in triggered events are paired with each other.  When both electrons are at midrapidity ($|\eta|<2.47$), the unlike-sign charged pairs with an invariant mass satisfying $66<m_{ee}<116$\,\GeV\ are accepted as signal $Z$ boson candidates.  The like-sign pairs in this window are used to estimate the combinatorial background, created primarily by jets. In total, 1647 unlike-sign pairs and 52 like-sign pairs are reconstructed.  The like-sign pairs are composed of combinatorial background and \Zee\ decays in which one of the electrons has misreconstructed charge.  The contribution from pairs with misreconstructed charge is estimated, using the MC simulation, to be half of the like-sign pairs, and the remainder is taken as an estimate of the background.  Pairs made of one midrapidity electron and one non-triggering forward-electron have a larger contribution from background and so an invariant mass window of $80<m_{ee}<100$\,\GeV\ is used to select $Z$ boson candidates.  To facilitate combination of all \Zee\ candidates, an acceptance correction is made to account for the smaller mass window.  No charge requirement is made for these candidates because the non-triggering electron is outside the acceptance of the ID and therefore does not have a reconstructed charge.  There are 264 such candidates, of which an estimated 5\% are background based on a fit of the invariant mass distribution.   The fit is performed in the range  $60<m_{ee}<120$\,\GeV\ using a signal shape from the MC simulation, and several background parametrizations assuming exponential or polynomial descriptions of the background. The mass distributions of \Zee\ candidates are shown in Figure \ref{fig:Zll_mass}(a) and  \ref{fig:Zll_mass}(b), along with the reconstructed MC simulation of the same quantity.  The estimated background is subtracted from the signal candidates differentially in rapidity, transverse momentum, and centrality.  
\begin{figure}[htb]
\begin{center}
\includegraphics[width=0.48\textwidth]{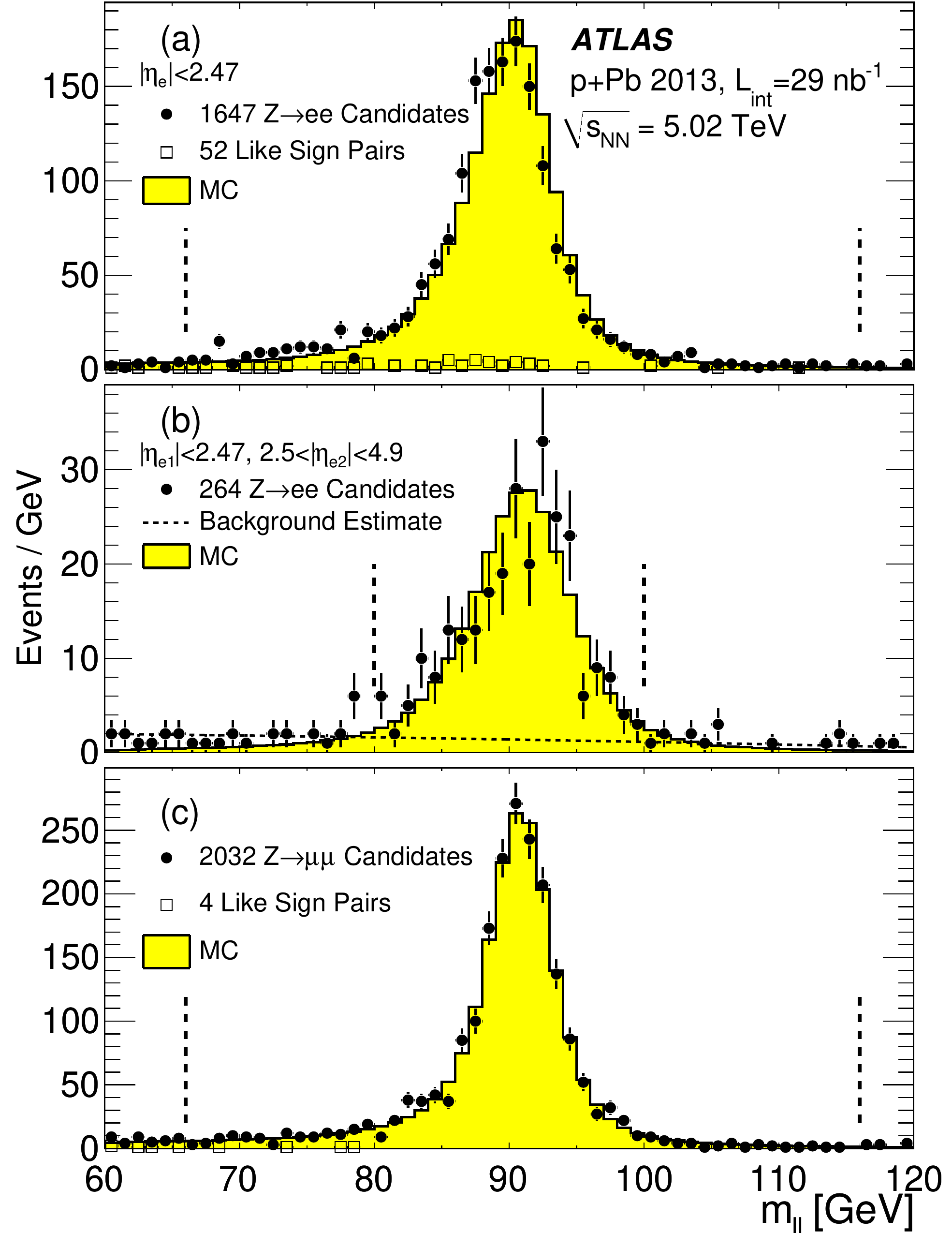} 
\caption[$Z$ bosons mass distribution.]{(Color online) The dilepton invariant mass distributions in data and MC simulation.  (a) \Zee\ candidates with both electrons at midrapidity ($|\eta|<2.47$); (b) candidates in which the non-triggering electron is a forward-electron ($2.5<|\eta|<4.9$); (c) \Zmm\ candidates.  The midrapidity \Zee\ and \Zmm\ distributions are overlaid with the mass distributions of like-sign lepton pairs.  For candidates in which the non-triggering electron is a forward-electron ($2.5<|\eta|<4.9$) the background, estimated based on a decomposition of the invariant mass distribution (see text), is shown.  In all plots the vertical lines indicate the mass window within which candidates are defined, and the MC simulation is normalized to the data inside this region.} 
\label{fig:Zll_mass}
\end{center}
\end{figure}

A similar procedure is also followed to select \Zmm\ candidates with an invariant mass of $66<m_{\mu\mu}<116$\,\GeV.  This selection yields 2032 unlike-sign charged candidates and 4 like-sign pairs; their mass distribution is shown in Figure \ref{fig:Zll_mass}(c).  
The MC simulation describes the data well in both lepton channels. The slight shift of the mass peak visible between the data and the simulation for dielectron events has only a very small effect on the calculation of corrections based on the MC simulation and is incorporated into the systematic uncertainty associated with electron reconstruction.  

Based on the like-sign pairs and MC simulation of charge misreconstruction, the uncertainty from the background subtraction is approximately 1\% in the \Zee\ channel for pairs in which both electrons are at midrapidity; in pairs involving a forward-electron the uncertainty ranges from 5\% to 20\% based on fits of the invariant mass distribution.  The background uncertainty in the \Zmm\ yield is negligible.  The largest source of correlated background in both lepton decay channels is the decay of $Z\rightarrow \tau\tau$ events into dielectron and dimuon pairs.  These are simulated and reconstructed just as \Zee\ and \Zmm\ are, but are found to have a negligible contribution following the analysis procedures. 

\subsection{Systematic uncertainties}
The dominant source of uncertainty in the \Zee\ measurement stems from imperfect knowledge of the efficiency of the electron identification requirements.  The uncertainty is driven by the limited number of events available for the tag-and-probe analysis which had to rely on 2013 \pPb\ collision data because the electron reconstruction performance changed due to detector conditions and occupancy compared to earlier \pp\ collision data.  The uncertainty is larger in pairs involving a forward-electron, and the sample has a lower purity than the sample of midrapidity electrons.   
Other electron uncertainties are significantly smaller and are associated with trigger efficiency, electron reconstruction efficiency and energy resolution, background subtraction (which becomes significant for forward-electrons), and charge misreconstruction.  In addition a small uncertainty stems from possible differences between the simulated \ystarZ\ distribution and the one measured in data.  The \Zee\ systematic uncertainties depend on  \ptZ, \ystarZ, and \pPb\ centrality and are summarized in Table \ref{table:zee_uncert}.

\begin{table}[htb]
\begin{center}
\begin{tabular}{c|c}
Source & Uncertainty Range [\%]  \\
\hline
Electron ID &  6--14 \\
Electron Reconstruction & 1--3\\
Electron Trigger   & 1--2\\
Background     & 1--3\\
MC \ystarZ\ Shape & 0--2\\
Forward-Electron Reconstruction &  4--15\\
Forward-Electron Background & 2--10\\
\end{tabular}
\caption{Relative systematic uncertainties, in percent, associated with the measurement of \Zee.  The uncertainties typically increase at the more forward rapidities.  Background includes charge misreconstruction, and electron reconstruction includes resolution.  The last two rows refer only to pairs where one of the electrons was reconstructed in the range $3.1<|\eta|<4.9$.  
}
\label{table:zee_uncert}
\end{center}
\end{table}

The conditions of muon reconstruction in the 2013 \pPb\ collision data closely resemble those in \pp\ collisions described in Ref.~\cite{muon_perf}.  The  small uncertainties from the more abundant \pp data are used in this analysis.  An uncertainty of 1\%, based on the performance of the muon reconstruction in high-pileup \pp collisions, is associated with the scale factors to account for possible differences between the datasets.  The uncertainties depend on \ptZ\ and \ystarZ.  Table \ref{table:zmm_uncert} summarizes the systematic uncertainties on the \Zmm\ measurement.
\begin{table}[htb]
\begin{center}
\begin{tabular}{c|c}
Source & Uncertainty Range [\%]  \\
\hline
Muon ID \& Reconstruction & 1.5 \\
Muon Trigger   & 1--2\\
Background     & $<$1\\
\pT\ Resolution &  $<$1\\
MC \ystarZ\ Shape & $<$1.5\\
\end{tabular}
\caption{Relative systematic uncertainties, in percent, associated with the measurement of \Zmm.   
}
\label{table:zmm_uncert}
\end{center}
\end{table}

In addition to the $Z$ boson measurement uncertainties,  a 2.7\% uncertainty is associated with the luminosity calculation.  For the centrality-dependent yields that are scaled by  \mTAA\ the uncertainties of the Glauber model calculations are taken from Ref.  \cite{pPb_multpaper}.  The correction to centrality due to bias from the presence of a hard scattering, taken from Ref.~\cite{dennis_peter_paper}, has uncertainties as shown in Figure \ref{fig:cent_corrections}.
\subsection{Lepton channel combination \label{sec:combination}}
The cross section in each leptonic decay channel is defined for the mass window $66<m_{Z}<116$ \GeV, the rapidity ranges $|\ystarZ|<$3.5 for \Zee\ and $-3<\ystarZ<2$ for \Zmm, and the full decay-lepton kinematic phase-space.  The \Zee\ and \Zmm\ yields, corrected for acceptance and efficiency, are used to calculate the cross section in each channel, and good agreement between the two is observed as is shown for the \ystarZ\ distributions in Figure \ref{fig:zll_comp}.  
\begin{figure}[htb]
\begin{center}
\includegraphics[width=0.48\textwidth]{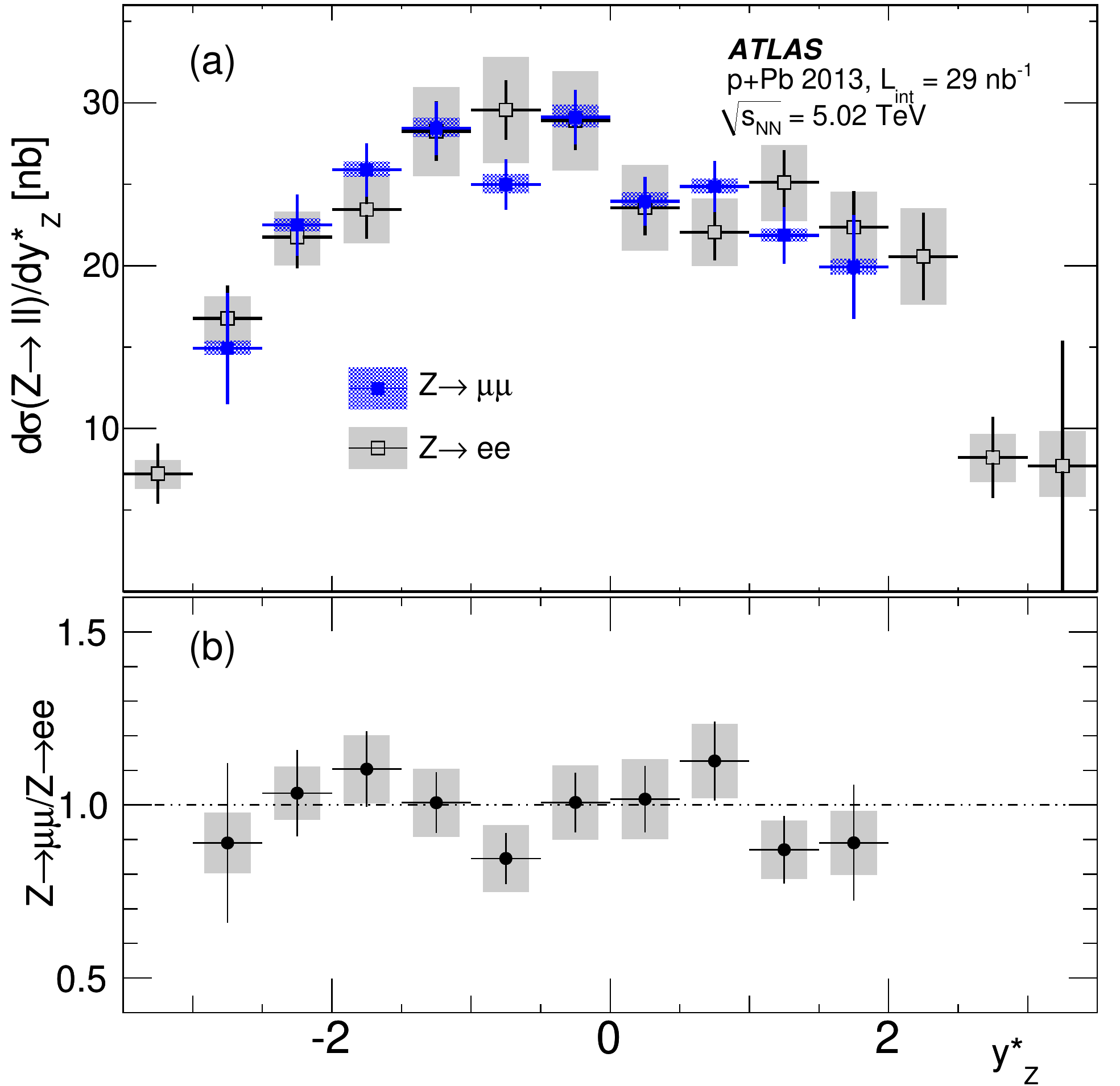}
\caption[Comparison of $d\sigma/d\ystarZ$ between \Zee~ and \Zmm.]{(Color online) (a) Differential $Z$ boson production cross section,$d\sigma/d\ystarZ$, as a function of $Z$ boson rapidity in the center-of-mass frame \ystarZ, for \Zee\ and \Zmm\ (upper panel).  (b) Their ratio.  Bars indicate statistical uncertainty and the shaded boxes systematic uncertainty.} 
\label{fig:zll_comp}
\end{center}
\end{figure}

The two decay channel results are combined to one set of \Zll\ data using the method described in Refs. \cite{hera_avg,hera_avg_formal}.
The technique uses a $\chi^{2}$ minimization procedure with a nuisance parameter formalism to combine the data sets coherently.  The procedure distinguishes between those systematic uncertainty sources that are uncorrelated bin-to-bin, uncorrelated across data sets, and fully correlated bin-to-bin and across data sets.  In this way, combined points are calculated to optimize the overall agreement of the data sets, given the correlation of the uncertainties.  This may result in differences in the combined \Zll\ data points relative to the \Zee\ data points in the rapidity regions in which there are no \Zmm\ data points.  Following this, an integrated cross section for the region $|\ystarZ|<3.5$ is defined for the combined \Zll\ points based on both the \Zee\ and \Zmm\ data even though the \Zmm\ data are limited to $-3<\ystarZ<2$.
The systematic uncertainties associated with the combined results are  
fully correlated bin-to-bin in each distribution.  They are approximately 3\% at midrapidity, and rise to about 10\% at forward and backward rapidity.

\section{Results}
\subsection{\Zll\ cross section}
From the combined \Zee\ and \Zmm\ data a total cross section of 139.8 $\pm$ 4.8 (statistical) $\pm$ 6.2 (systematic)  $\pm$ 3.8 (luminosity) nb is obtained in the $|\ystarZ|<3.5$ acceptance.  Based on the MC simulation (and the models discussed below) this acceptance covers approximately 99.5\% of the total \Zll\ cross section.  Restricting the results to the smaller rapidity interval of $-3<\ystarZ<2$, the cross section is 119.3 $\pm$ 2.2 (stat.) $\pm$ 3.4 (syst.)  $\pm$ 3.2 (lumi.) nb.  Table \ref{table:comb_cross} lists the integrated cross section in the larger and smaller rapidity ranges as measured for each channel and their combination.

\begin{table*}[htb]
\begin{center}
\begin{tabular}{lcccc}
\ystarZ\ &$[-2,0]$&$[0,2]$&$[-3,2]$&$[-3.5,3.5]$\\
\hline
\Zmm& $54.2\pm1.6\pm1.3$ & $45.3\pm2.1\pm0.9$ & $118.2\pm3.3 \pm 2.6$ &N/A\\ 
\Zee& $55.1\pm 1.8 \pm5.9$ & $46.5\pm2.2\pm5.0$ & $121\pm3\pm13$ & $143\pm5\pm17$ \\ 
\Zll  & $54.4\pm 1.3 \pm1.4$ & $45.9 \pm 1.4 \pm 1.4$ &$119.3\pm 2.2\pm3.4$ &$139.8\pm4.8 \pm6.2$ \\      
\hline
CT10 (NLO) & $47.4\pm0.9$ &  $46.8\pm0.9$ &$110.8\pm2.9$ & $132.2\pm3.3$ \\
CT10+EPS09 (NLO) & $48.7\pm1.0$ &  $43.5\pm1.1$ &$108.6\pm3.1$ & $127.4\pm3.6$ \\
MSTW2008 (NNLO) & $48.3 ^{+1.2}_{-0.9}$ & $47.9 ^{+1.2}_{-0.9}$ &$113.5 ^{+2.8}_{-2.2}$ & $135.2 ^{+3.4}_{-2.7}$ \\
\end{tabular}
\caption{The measured integrated cross section, in nb, for several rapidity ranges, for \Zmm, \Zee, and the combined \Zll.  The first uncertainty listed is statistical and the second systematic.  There is an additional 2.7\% luminosity uncertainty on each cross section.  The cross sections predicted by the models (see text) are also shown.  The uncertainties listed with the model calculations are the PDF and scale uncertainties added in quadrature.}
\label{table:comb_cross}
\end{center}
\end{table*}

The measured cross section may be compared to a \pPb\ model prediction composed of a linear sum of the nucleon-nucleon cross sections: 82$\sigma(\pp\rightarrow Z+X)$ + 126$\sigma(\pn\rightarrow Z+X)$, corresponding to the numbers of protons and neutrons in the Pb ion.  The value of $\sigma(\pn\rightarrow Z+X)$ is 2\% higher than that of $\sigma(\pp\rightarrow Z+X)$ in all models discussed below. Calculating the baseline nucleon-nucleon cross sections using the CT10 PDF at next-to-leading order (NLO), as in the corresponding MC simulation, the model yields values of 132.2 $\pm$ 3.3 nb in the range  $|\ystarZ|<3.5$, and 110.8 $\pm$ 2.9 nb for $-3<\ystarZ<2$ where the uncertainties are the sums in quadrature of PDF and scale (renormalization and factorization) uncertainties.  Using the MSTW2008 PDF, calculated with {\scshape Fewz} \cite{FEWZ_paper} at next-to-next-to-leading order (NNLO), cross sections of 135.2 $^{+3.4}_{-2.7}$ nb are obtained for  $|\ystarZ|<3.5$ and 113.5 $^{+2.8}_{-2.2}$ nb for $-3<\ystarZ<2$.  At NLO the results from MSTW2008 are very close to the CT10 results.  In addition to the simple model of the \pPb\ $Z$ boson cross section as a linear sum of nucleon-nucleon cross sections, calculations are performed incorporating nuclear corrections of the PDF.  Including the EPS09 modifications \cite{hannu} to the CT10 PDF results in cross sections of 127.4 $\pm$ 3.6 nb and 108.6 $\pm$ 3.1 nb, respectively.

For a more detailed understanding of $Z$ boson production, the measured cross section as a function of the $Z$ boson rapidity is presented in Figure \ref{fig:zll_dsig_drap} and compared to model calculations.  The data are seen to be strongly asymmetric about \ystarZ=0.  
   The CT10+EPS09 calculations come closest to reproducing the shape of the measured \ystarZ\ differential cross section.  A $\chi^{2}$ test of compatibility between the data and the model shapes (irrespective of normalization) finds that the CT10+EPS09 shape of the \ystarZ\ distribution gives a $p$-value of 0.79.   The unmodified CT10 calculation and MSTW2008 calculations have $p$-values of 0.07 and 0.01, respectively.  A Kolmogorov-Smirnov test was also performed and resulted in probabilities of 0.96, 0.09, and 0.07 for CT10+EPS09, CT10, and MSTW2008 model calculations.  This is consistent with the preference for the observation of nuclear correction effects as in the $\chi^{2}$ test.

\begin{figure}[htb]
\begin{center}
\includegraphics[width=0.48\textwidth]{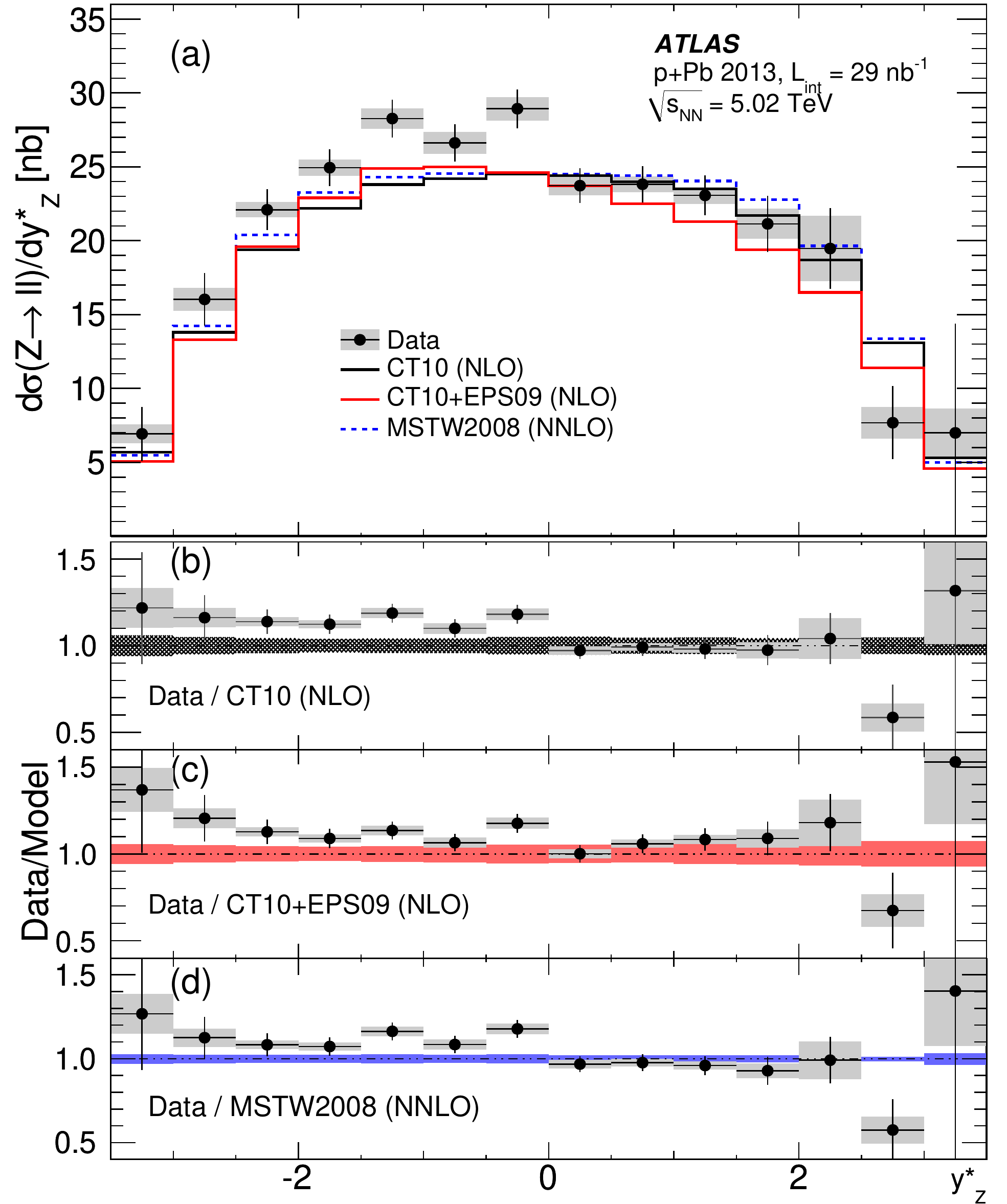}
\caption{(Color online) (a) The $d\sigma/d\ystarZ$ distribution from \Zll, shown along with several model calculations in the upper panel.  The bars indicate statistical uncertainty and the shaded boxes systematic uncertainty on the data; the uncertainties on the model calculations are not shown.  (b-d) Ratios of the data to the models.  The uncertainties of the model calculations (scale and PDF uncertainties added in quadrature) are shown as bands around unity in each panel. An additional 2.7\% luminosity uncertainty on the cross section is not shown.}
\label{fig:zll_dsig_drap}
\end{center}
\end{figure}

Nuclear modification of PDFs is fundamentally related to the Bjorken $x$ of the relevant parton.  At leading order, $x_{p}$ in the proton and $x_{Pb}$ in the lead nucleus are related to the reconstructed $Z$ boson kinematics by:
\begin{equation}
x_{p}= \frac{m_{\ell\ell}e^{\ystarZ}}{\sqrt{s_\mathrm{NN}}}, \hspace{0.6cm} x_{Pb}=\frac{m_{\ell\ell}e^{-\ystarZ}}{\sqrt{s_\mathrm{NN}}}  ~. 
\label{eq:bjorkenx}
\end{equation}
The resulting $x_{Pb}$ distribution is shown in Figure \ref{fig:zll_dsig_dx2} and compared to model calculations.
\begin{figure}[htb]
\begin{center}
\includegraphics[width=0.48\textwidth]{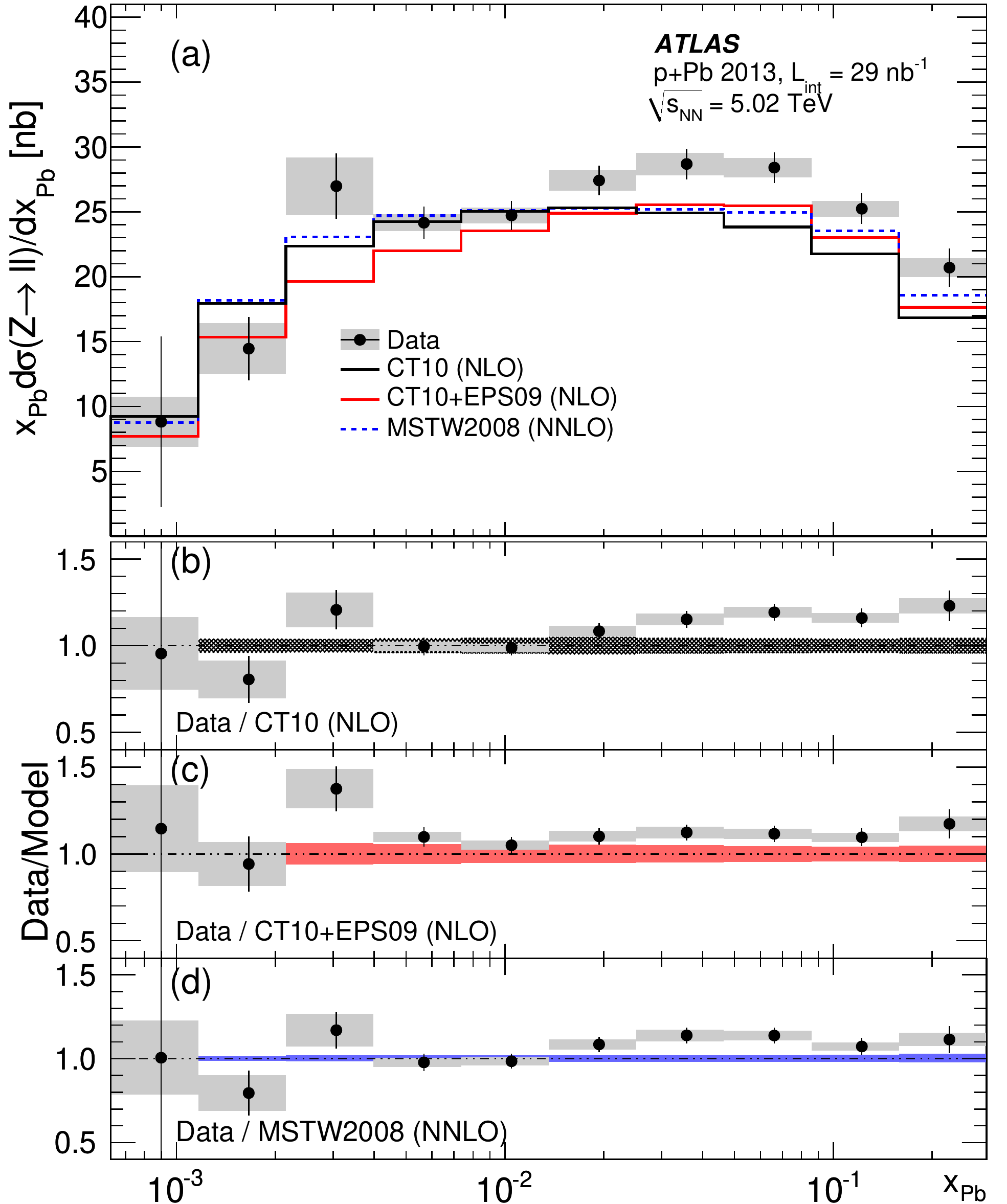}
\caption{(Color online) (a) The differential cross section of $Z$ boson production multiplied by the Bjorken $x$ of the parton in the lead nucleus, $x_{Pb} d\sigma /dx_{Pb}$, as a function of $x_{Pb}$ using \Zll\ events shown along with several model calculations in the upper panel.  The bars indicate statistical uncertainty and the shaded boxes systematic uncertainty on the data; the uncertainties on the model calculations are not shown.  (b-d) Ratios of the data to the models. The uncertainties of the model calculations are shown as bands around unity in each panel.  There is an additional 2.7\% luminosity uncertainty on the cross section.} 
\label{fig:zll_dsig_dx2}
\end{center}
\end{figure}

Figure \ref{fig:zll_dsig_dpt} shows the \ptZ\ distributions for $-3<\ystarZ<2$, 
and separately for $-2<\ystarZ<0$ and $0<\ystarZ<2$.
These are compared to the baseline CT10 model.  The \ptZ\ dependence is less sensitive to nuclear effects and good agreement between the experimental measurement and the MC simulation shape is observed.
\begin{figure}[htb]
\begin{center}
\includegraphics[width=0.42\textwidth]{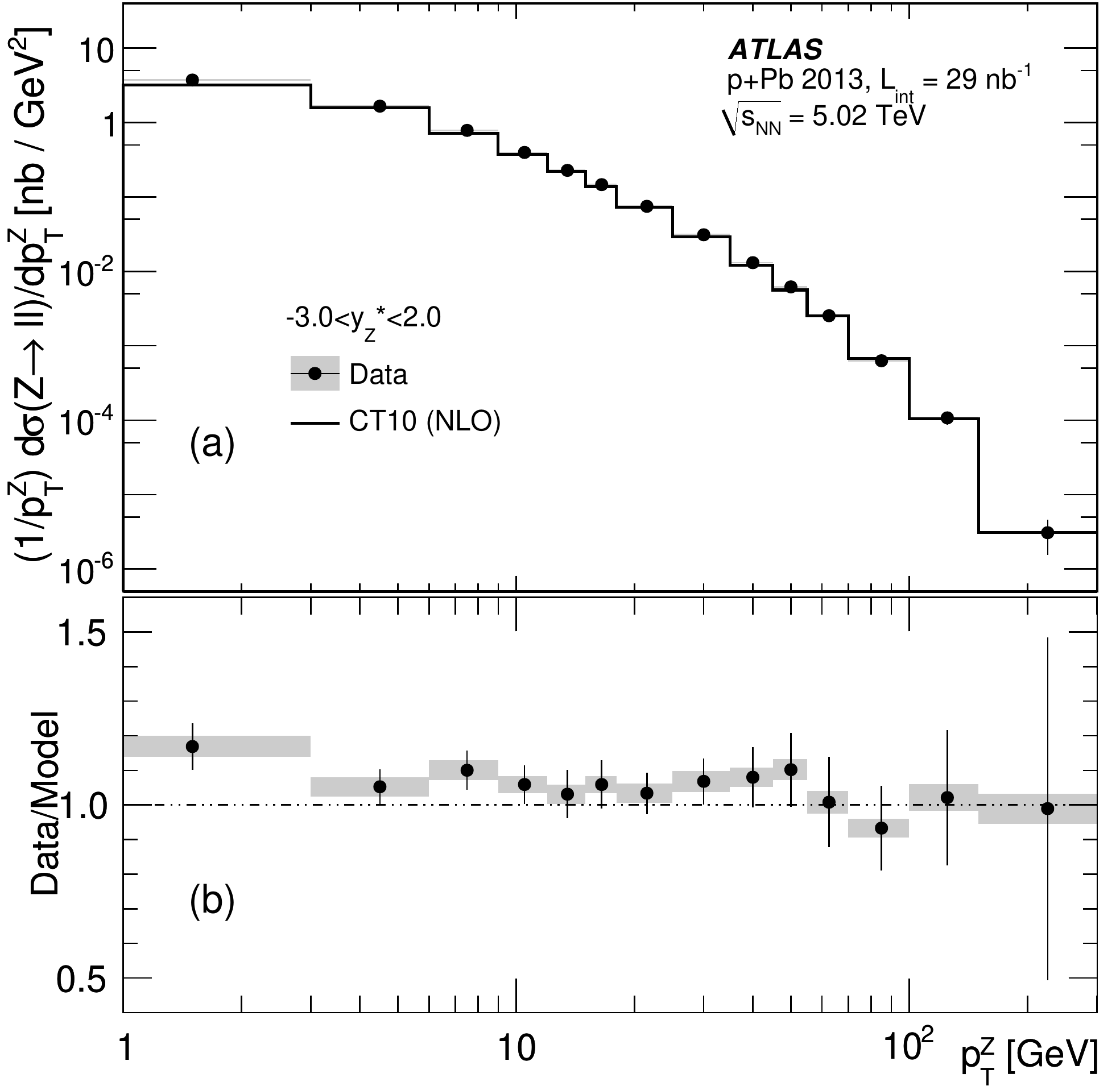}
\includegraphics[width=0.42\textwidth]{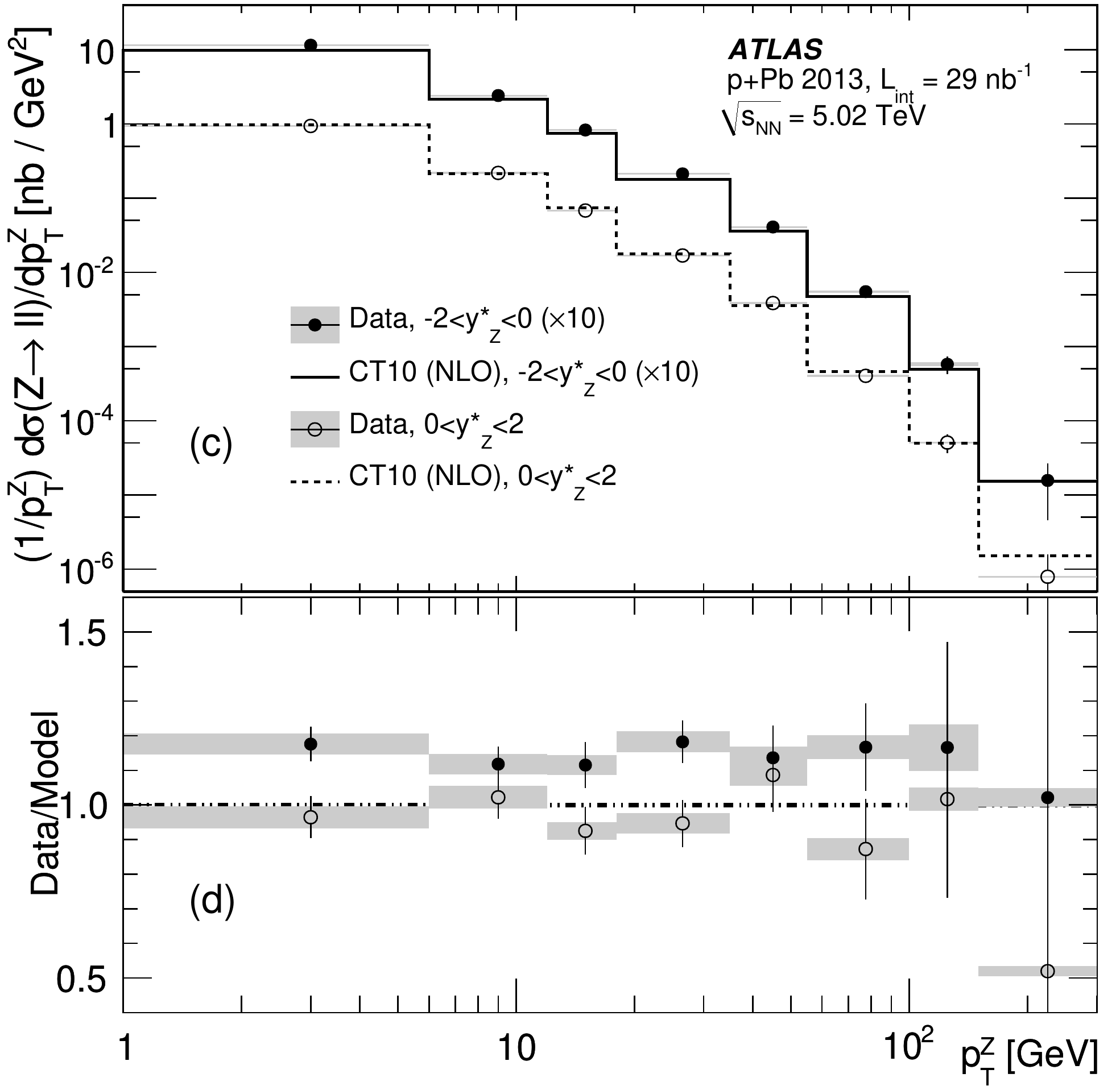}
\caption{(Color online) (a, c) The distributions of the differential cross section of $Z$
boson production as a function of the transverse momentum of the $Z$, \ptZ,  shown along with the CT10 model calculation. (b, d) Ratio of the data to the model.  (a,b) for $-3<\ystarZ<2$; (c,d) for $-2<\ystarZ<0$ and $0<\ystarZ<2$.  The bars indicate statistical uncertainty and the shaded boxes systematic uncertainty.  The leftmost bin represents the range 0--3 GeV.  An additional 2.7\% luminosity uncertainty of the cross section is not shown.
} 
\label{fig:zll_dsig_dpt}
\end{center}
\end{figure}

\subsection{Centrality-dependent yield \label{sec:cent_yields}}
Results are presented for the centrality-dependent $Z$ boson yield.  If the rate of $Z$ boson production were consistent with geometric expectations, then the $Z$ boson yield divided by \avgNcoll should be  independent of centrality.  To investigate this, the yield of $Z$ bosons per event scaled by \avgNcoll, within  $-3<\ystarZ<2$, is displayed as a function of \avgNpart in Figure \ref{fig:centrality_yields}.  The yield is independent of centrality defined using the standard Glauber model.    
Using the GGCF centrality models increases \avgNcoll\ in central events and reduces it in peripheral events, consequently the yield divided by \avgNcoll\ is reduced in central events and increased in peripheral events.  Figure \ref{fig:centrality_yields} also shows the yield without the application of the centrality bias corrections discussed in Sec.~\ref{sec:centrality}. 
\begin{figure}[htb]
\begin{center}
\includegraphics[width=0.48\textwidth]{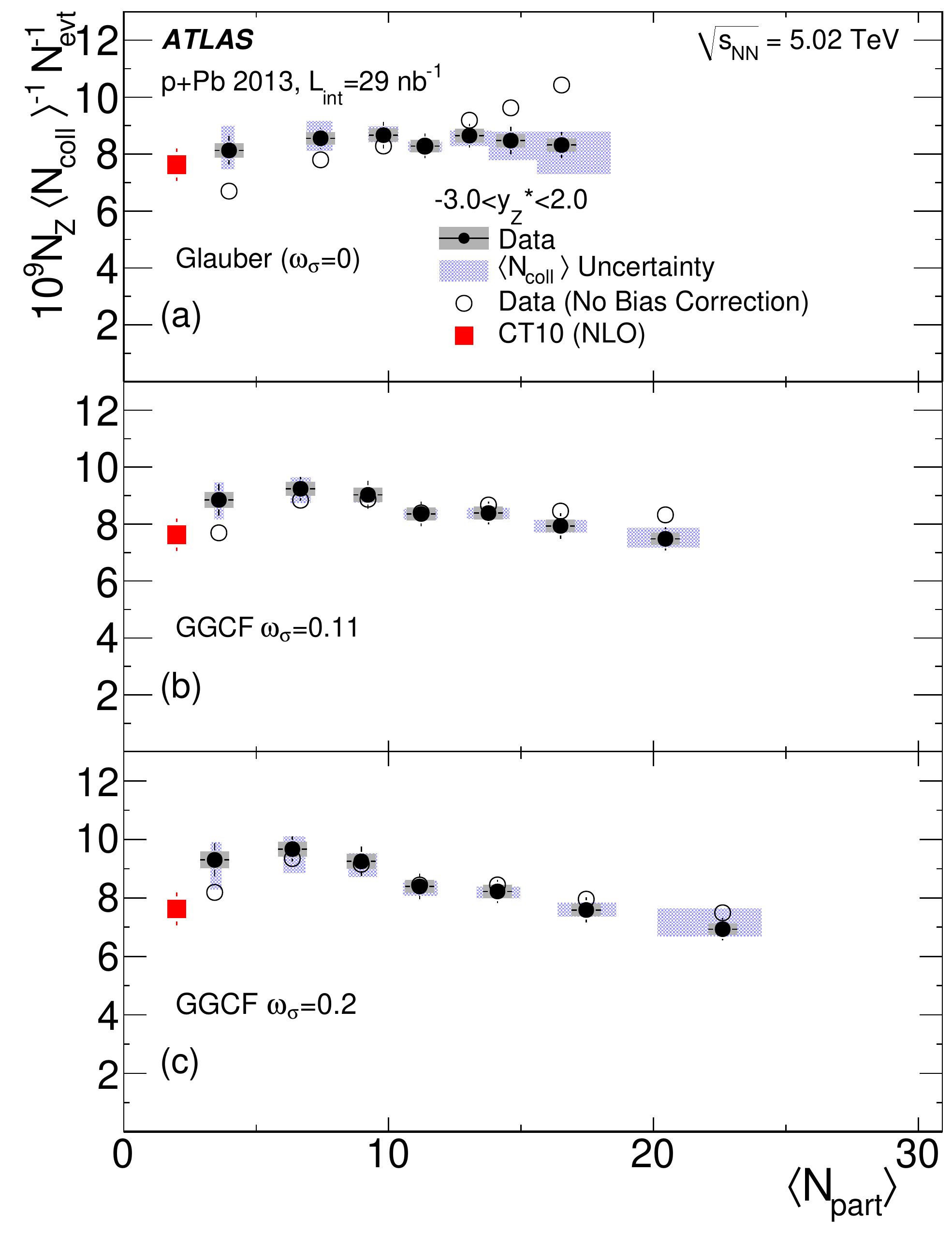}
\caption{(Color online)  The yield of $Z$ bosons per event scaled by the mean number of
nucleon-nucleon collisions \avgNcoll as a function of the mean number of participating nucleons  \avgNpart.  
Each panel uses a different Glauber model configuration in calculating \avgNpart (and \avgNcoll).  (a) Standard Glauber model with no Glauber-Gribov color fluctuations; (b) GGCF with $\omega_{\sigma} = 0.11$; (c) GGCF with $\omega_{\sigma} = 0.2$.  The data are compared to the CT10 model prediction plotted at \avgNpart = 2.  The bars indicate statistical uncertainty and the shaded boxes systematic uncertainty.  The systematic uncertainties are correlated bin-by-bin.  The \avgNcoll uncertainty plotted does not include the bin-by-bin fully correlated uncertainty stemming from the uncertainty on $\sigma(N+N\rightarrow X)$, which is instead included in the CT10 prediction uncertainty.  As a reference, the data are plotted as they would be with no centrality bias correction in the open points.}
\label{fig:centrality_yields}
\end{center}
\end{figure}

The ATLAS Collaboration has previously measured the inclusive charged-hadron multiplicity in \pPb\ collisions as a function of centrality \cite{pPb_multpaper}, and the centrality dependence of that quantity is similar to that observed in the present measurement.  In order to quantify the similarity, the ratio $(dN_{Z}/d\ystarZ)/(dN_\mathrm{ch}/d\eta)$ is plotted vs \avgNpart in Figure \ref{fig:zcharge_ratio}.  The charged-particle yield is expected to scale with \avgNpart and the $Z$ boson yield with $\avgNcoll=\avgNpart-1$, and so the ratio is fit to a function with the form $a\cdot(\avgNpart-1)/\avgNpart$.
This function describes the data well for the GGCF cases, and less so for the standard Glauber model.  

\begin{figure}[htb]
\begin{center}
\includegraphics[width=0.48\textwidth]{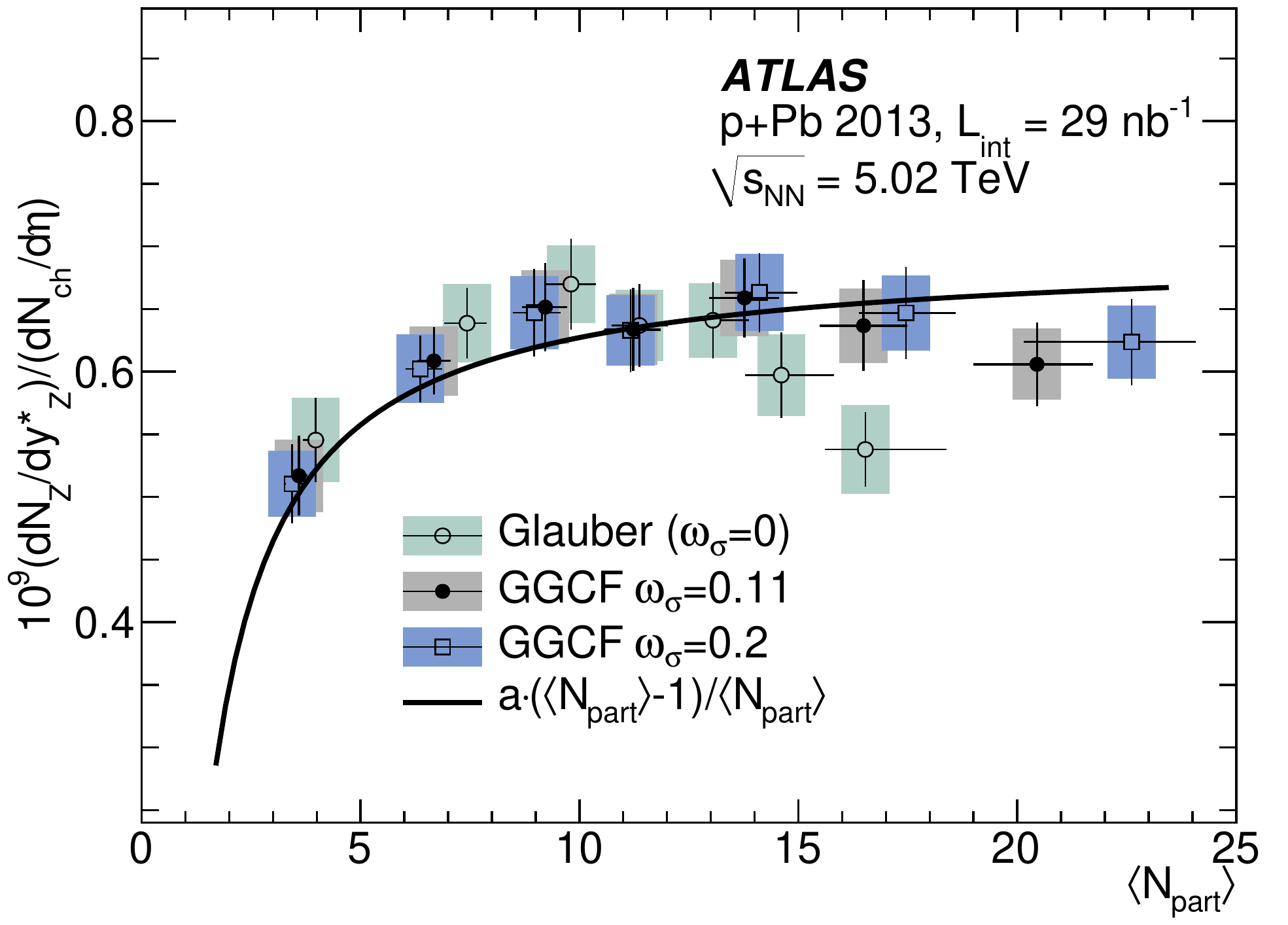}
\caption{(Color online) The ratio of the $Z$ boson multiplicity to the inclusive charged particle multiplicity, $(dN_{Z}/d\ystarZ)/(dN_\mathrm{ch}/d\eta)$, as a function of \avgNpart.  A function of the form $a\cdot(\avgNpart-1)/\avgNpart$ is also shown.  The normalization $a$ is set based on the GGCF with $\omega_{\sigma} = 0.11$ points.  Statistical uncertainties are plotted as bars and systematic uncertainties as shaded boxes.
} 
\label{fig:zcharge_ratio}
\end{center}
\end{figure}

To further investigate the behavior observed in the rapidity differential cross section, the \ystarZ\ dependence of the Z boson yield in different centrality bins is also measured, as shown in Figure \ref{fig:centbinned_ydists}.  The differences between data and the model are larger in central collisions.  The \avgNcoll-scaled ratio of central to peripheral data, \RCP, defined as 
\begin{equation}
\RCP(\ystarZ)=\frac{\avgNcoll^\mathrm{peripheral}}{\avgNcoll^\mathrm{central}}\times \frac{dN_{Z}^\mathrm{central}/d\ystarZ}{dN_{Z}^\mathrm{peripheral}/d\ystarZ}   
\label{eq:rcp}
\end{equation}
is used to observe changes in the rapiditiy distribution for different centrality bins in a model-independent way and is shown in Figure \ref{fig:centbinned_ydists}.
\begin{figure}[htb]
\begin{center}
\includegraphics[width=0.48\textwidth]{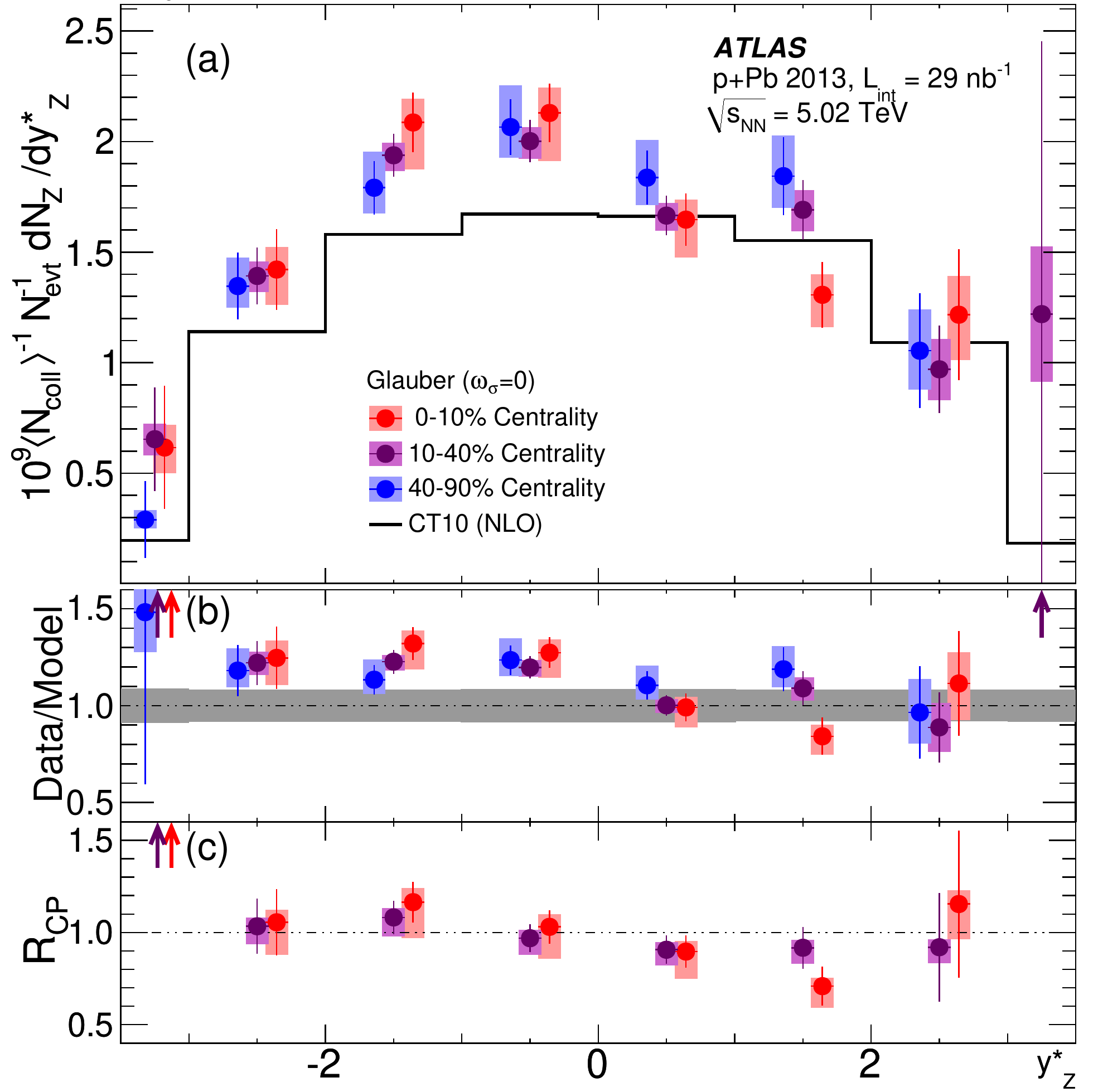}
\caption{(Color online) (a) The rapidity differential $Z$ boson yields, scaled by \avgNcoll, for three centrality ranges compared with the CT10 model calculation.  The bars indicate statistical uncertainty and the shaded boxes systematic uncertainty.  The \avgNcoll is defined with the standard Glauber model ($\omega_{\sigma}$=0). The scale uncertainty stemming from the centrality calculation for each bin is included in the systematic uncertainty.  The uncertainty associated with the model is not plotted.  (b) The ratios of the data to the model.  The uncertainty of the model added in quadrature with the scale uncertainty due to uncertainty in the inclusive $NN$ cross section is shown as a band around unity.  (c)  \RCP (see text for details). The 0--10\% and 40--90\% centrality points are offset for visual clarity.  The arrows in the lower panels indicate values outside the plotted axes.} 
\label{fig:centbinned_ydists}
\end{center}
\end{figure}
Events with 40--90\% centrality define the peripheral event selection, and two central selections, 0--10\% and 10--40\%, are compared with it.  
A linear fit of the \RCP(\ystarZ) for 0--10\% centrality, results in a slope of $-0.11 \pm 0.04$, which suggests that the \ystarZ\ distribution may be different in the most central events compared to peripheral events.  For 10--40\% centrality, the slope is $-0.05 \pm 0.03$.
\FloatBarrier

\section{Summary}
The $Z$ boson production cross section has been measured in \pPb collisions at \energy\ with the ATLAS detector at the LHC, using \Zee\ (\Zmm) decays in a 29.4 \inb\ (28.1 \inb) data sample.  It is found to be slightly higher than predictions based on perturbative QCD calculations.  Disregarding the difference in overall normalization, the shapes of the \ystarZ\- and $x_\mathrm{Pb}$- dependent cross sections are somewhat better described by models that include nuclear modification of the lead nucleus PDF compared to those that do not, although models without nuclear modification are not excluded.  Following the application of a centrality bias correction, the centrality-dependent yield is found to scale with \avgNcoll.  In addition, the centrality dependence of the \ystarZ\ distribution was  studied, and the asymmetry in \ystarZ\ was found to be slightly larger in more central events.  Integrated over \ystarZ, the centrality dependence appears to be consistent with binary scaling and is similar to the production of inclusive charged particles.

\section*{Acknowledgments}


We thank CERN for the very successful operation of the LHC, as well as the
support staff from our institutions without whom ATLAS could not be
operated efficiently.

We acknowledge the support of ANPCyT, Argentina; YerPhI, Armenia; ARC,
Australia; BMWFW and FWF, Austria; ANAS, Azerbaijan; SSTC, Belarus; CNPq and FAPESP,
Brazil; NSERC, NRC and CFI, Canada; CERN; CONICYT, Chile; CAS, MOST and NSFC,
China; COLCIENCIAS, Colombia; MSMT CR, MPO CR and VSC CR, Czech Republic;
DNRF, DNSRC and Lundbeck Foundation, Denmark; EPLANET, ERC and NSRF, European Union;
IN2P3-CNRS, CEA-DSM/IRFU, France; GNSF, Georgia; BMBF, DFG, HGF, MPG and AvH
Foundation, Germany; GSRT and NSRF, Greece; RGC, Hong Kong SAR, China; ISF, MINERVA, GIF, I-CORE and Benoziyo Center, Israel; INFN, Italy; MEXT and JSPS, Japan; CNRST, Morocco; FOM and NWO, Netherlands; BRF and RCN, Norway; MNiSW and NCN, Poland; GRICES and FCT, Portugal; MNE/IFA, Romania; MES of Russia and NRC KI, Russian Federation; JINR; MSTD,
Serbia; MSSR, Slovakia; ARRS and MIZ\v{S}, Slovenia; DST/NRF, South Africa;
MINECO, Spain; SRC and Wallenberg Foundation, Sweden; SER, SNSF and Cantons of
Bern and Geneva, Switzerland; NSC, Taiwan; TAEK, Turkey; STFC, the Royal
Society and Leverhulme Trust, United Kingdom; DOE and NSF, United States of
America.

The crucial computing support from all WLCG partners is acknowledged
gratefully, in particular from CERN and the ATLAS Tier-1 facilities at
TRIUMF (Canada), NDGF (Denmark, Norway, Sweden), CC-IN2P3 (France),
KIT/GridKA (Germany), INFN-CNAF (Italy), NL-T1 (Netherlands), PIC (Spain),
ASGC (Taiwan), RAL (UK) and BNL (USA) and in the Tier-2 facilities
worldwide.

\bibliographystyle{atlasBibStyleWoTitle}
\bibliography{pPb_zll_paper.bib}

\providecommand{\href}[2]{#2}\begingroup\raggedright\begin{thebibliography}{10}

\bibitem{cms_photons}
{CMS Collaboration},
\href{http://dx.doi.org/10.1016/j.physletb.2012.02.077}{Phys. Lett. {\bfseries
  B710} (2012) 256--277}.

\bibitem{cms_ws}
{CMS Collaboration},
\href{http://dx.doi.org/10.1016/j.physletb.2012.07.025}{Phys. Lett. {\bfseries
  B715} (2012) 66--87}.

\bibitem{wboson_paper}
{ATLAS Collaboration},
\href{http://dx.doi.org/10.1140/epjc/s10052-014-3231-6}{Eur. Phys. J.
  {\bfseries C75} no.~1, (2015) 23}.

\bibitem{PhysRevLett.110.022301}
{ATLAS Collaboration},
\href{http://dx.doi.org/10.1103/PhysRevLett.110.022301}{Phys. Rev. Lett.
  {\bfseries 110} no.~2, (2013) 022301}.

\bibitem{cms_z}
{CMS Collaboration},
\href{http://dx.doi.org/10.1007/JHEP03(2015)022}{JHEP {\bfseries 1503} (2015)
  022}.

\bibitem{Sjostrand:2007gs}
T.~Sj{\"o}strand, S.~Mrenna,  and P.~Z. Skands,
\href{http://dx.doi.org/10.1016/j.cpc.2008.01.036}{Comput. Phys. Commun.
  {\bfseries 178} (2008) 852--867}.

\bibitem{hannuDIS}
H.~Paukkunen, PoS {\bfseries DIS2014} (2014) 053,
\href{http://arxiv.org/abs/1408.4657}{{\ttfamily arXiv:1408.4657 [hep-ph]}}.

\bibitem{alice_pPb_flow1}
{ALICE} Collaboration, B.~Abelev {et~al.},
\href{http://dx.doi.org/10.1016/j.physletb.2013.01.012}{Phys. Lett. {\bfseries
  B719} (2013) 29--41}.

\bibitem{cms_pPb_flow1}
{CMS Collaboration},
\href{http://dx.doi.org/10.1016/j.physletb.2012.11.025}{Phys. Lett. {\bfseries
  B718} (2013) 795--814}.

\bibitem{atlas_pPb_flow1}
{ATLAS Collaboration},
\href{http://dx.doi.org/10.1103/PhysRevLett.110.182302}{Phys. Rev. Lett.
  {\bfseries 110} no.~18, (2013) 182302}.

\bibitem{alice_pPb_flow2}
{ALICE} Collaboration, B.~Abelev {et~al.},
\href{http://dx.doi.org/10.1016/j.physletb.2013.08.024}{Phys. Lett. {\bfseries
  B726} (2013) 164--177}.

\bibitem{cms_pPb_flow2}
{CMS Collaboration},
\href{http://dx.doi.org/10.1016/j.physletb.2013.06.028}{Phys. Lett. {\bfseries
  B724} (2013) 213--240}.

\bibitem{atlas_pPb_flow2}
{ATLAS Collaboration},
\href{http://dx.doi.org/10.1016/j.physletb.2013.06.057}{Phys. Lett. {\bfseries
  B725} (2013) 60--78}.

\bibitem{atlas_pPb_ridge}
{ATLAS Collaboration},
\href{http://dx.doi.org/10.1103/PhysRevC.90.044906}{Phys. Rev. {\bfseries C90}
  no.~4, (2014) 044906}.

\bibitem{ATLAS:2014cpa}
{ATLAS Collaboration},
\href{http://dx.doi.org/10.1016/j.physletb.2015.07.023}{Phys. Lett. {\bfseries
  B748} (2015) 392--413}.

\bibitem{CMS_pPbdijet_paper}
{CMS Collaboration},
\href{http://dx.doi.org/10.1140/epjc/s10052-014-2951-y}{Eur. Phys. J.
  {\bfseries C74} no.~7, (2014) 2951}.

\bibitem{CMS_W_pPb}
{CMS Collaboration},
\href{http://dx.doi.org/10.1016/j.physletb.2015.09.057}{Phys. Lett. {\bfseries
  B750} (2015) 565--586}.

\bibitem{lhcb_Z_pPb}
{LHCb} Collaboration, R.~Aaij {et~al.},
\href{http://dx.doi.org/10.1007/JHEP09(2014)030}{JHEP {\bfseries 1409} (2014)
  030}.

\bibitem{Aad:2008zzm}
{ATLAS Collaboration},
\href{http://dx.doi.org/10.1140/epjc/s10052-013-2518-3}{Eur. Phys. J.
  {\bfseries C73} no.~8, (2013) 2518}.

\bibitem{Aad:2012xs}
{ATLAS Collaboration},
\href{http://dx.doi.org/10.1140/epjc/s10052-011-1849-1}{Eur. Phys. J.
  {\bfseries C72} (2012) 1849}.

\bibitem{lumi_paper}
{ATLAS Collaboration},
  \href{http://dx.doi.org/10.1140/epjc/s10052-013-2518-3}{Eur. Phys. J.
  {\bfseries C73} (2013) 1--39}.

\bibitem{pPb_multpaper}
{ATLAS Collaboration},
\href{http://arxiv.org/abs/1508.00848}{{\ttfamily arXiv:1508.00848 [hep-ex]}}.

\bibitem{eperf_2014paper}
{ATLAS Collaboration},
\href{http://dx.doi.org/10.1140/epjc/s10052-014-2941-0}{Eur. Phys. J.
  {\bfseries C74} no.~7, (2014) 2941}.

\bibitem{muon_perf}
{ATLAS Collaboration},
\href{http://dx.doi.org/10.1140/epjc/s10052-014-3130-x}{Eur. Phys. J.
  {\bfseries C74} no.~11, (2014) 3130}.

\bibitem{Glauber}
M.~L. Miller, K.~Reygers, S.~J. Sanders,  and P.~Steinberg,
\href{http://dx.doi.org/10.1146/annurev.nucl.57.090506.123020}{Ann.Rev.Nucl.Part.Sci.
  {\bfseries 57} (2007) 205--243}.

\bibitem{gribov1}
V.~Guzey and M.~Strikman,
\href{http://dx.doi.org/10.1016/j.physletb.2005.11.065,
  10.1016/j.physletb.2008.04.010}{Phys. Lett. {\bfseries B633} (2006)
  245--252}.

\bibitem{gribov2}
M.~Alvioli and M.~Strikman,
\href{http://dx.doi.org/10.1016/j.physletb.2013.04.042}{Phys. Lett. {\bfseries
  B722} (2013) 347--354}.

\bibitem{PHENIX_dAu_cent}
{PHENIX} Collaboration, A.~Adare {et~al.},
\href{http://dx.doi.org/10.1103/PhysRevC.90.034902}{Phys. Rev. {\bfseries C90}
  no.~3, (2014) 034902}.

\bibitem{ALICE_pPb_cent}
{ALICE} Collaboration, J.~Adam {et~al.},
\href{http://dx.doi.org/10.1103/PhysRevC.91.064905}{Phys. Rev. {\bfseries C91}
  no.~6, (2015) 064905}.

\bibitem{dennis_peter_paper}
D.~V. Perepelitsa and P.~A. Steinberg,
\href{http://arxiv.org/abs/1412.0976}{{\ttfamily arXiv:1412.0976 [nucl-ex]}}.

\bibitem{powhegz}
S.~Alioli, P.~Nason, C.~Oleari,  and E.~Re,
\href{http://dx.doi.org/10.1088/1126-6708/2008/07/060}{JHEP {\bfseries 0807}
  (2008) 060}.

\bibitem{CT10_paper}
H.-L. Lai, M.~Guzzi, J.~Huston, Z.~Li, P.~M. Nadolsky, J.~Pumplin,  and C.-P.
  Yuan,
\href{http://dx.doi.org/10.1103/PhysRevD.82.074024}{Phys. Rev. {\bfseries D82}
  (2010) 074024}.

\bibitem{GEANT4_paper}
{GEANT4} Collaboration, S.~Agostinelli {et~al.},
  \href{http://dx.doi.org/10.1016/S0168-9002(03)01368-8}{Nucl. Instrum. Meth.
  {\bfseries A506} (2003) 250--303}.

\bibitem{ATLAS_sim}
{ATLAS Collaboration},
\href{http://dx.doi.org/10.1140/epjc/s10052-010-1429-9}{Eur. Phys. J.
  {\bfseries C70} (2010) 823--874}.

\bibitem{hera_avg}
A.~Glazov, \href{http://dx.doi.org/10.1063/1.2122026}{AIP Conference
  Proceedings {\bfseries 792} (2005) 237--240}.

\bibitem{hera_avg_formal}
{H1} Collaboration, F.~Aaron {et~al.},
\href{http://dx.doi.org/10.1140/epjc/s10052-009-1128-6}{Eur. Phys. J.
  {\bfseries C63} (2009) 625--678}.

\bibitem{FEWZ_paper}
R.~Gavin, Y.~Li, F.~Petriello,  and S.~Quackenbush,
\href{http://dx.doi.org/10.1016/j.cpc.2011.06.008}{Comput. Phys. Commun.
  {\bfseries 182} (2011) 2388--2403}.

\bibitem{hannu}
H.~Paukkunen and C.~A. Salgado,
\href{http://dx.doi.org/10.1007/JHEP03(2011)071}{JHEP {\bfseries 1103} (2011)
  071}.

\end{thebibliography}\endgroup

\newpage 
\begin{flushleft}
{\Large The ATLAS Collaboration}

\bigskip

G.~Aad$^{\rm 85}$,
B.~Abbott$^{\rm 113}$,
J.~Abdallah$^{\rm 151}$,
O.~Abdinov$^{\rm 11}$,
R.~Aben$^{\rm 107}$,
M.~Abolins$^{\rm 90}$,
O.S.~AbouZeid$^{\rm 158}$,
H.~Abramowicz$^{\rm 153}$,
H.~Abreu$^{\rm 152}$,
R.~Abreu$^{\rm 116}$
% ...
\bigskip
\\
% ...
$^{11}$ Institute of Physics, Azerbaijan Academy of Sciences, Baku, Azerbaijan\\
% ...
$^{85}$ CPPM, Aix-Marseille Universit{\'e} and CNRS/IN2P3, Marseille, France\\
% ...
$^{90}$ Department of Physics and Astronomy, Michigan State University, East Lansing MI, United States of America\\
% ...
$^{107}$ Nikhef National Institute for Subatomic Physics and University of Amsterdam, Amsterdam, Netherlands\\
% ...
$^{113}$ Homer L. Dodge Department of Physics and Astronomy, University of Oklahoma, Norman OK, United States of America\\
% ...
$^{116}$ Center for High Energy Physics, University of Oregon, Eugene OR, United States of America\\
% ...
$^{151}$ Institute of Physics, Academia Sinica, Taipei, Taiwan\\
$^{152}$ Department of Physics, Technion: Israel Institute of Technology, Haifa, Israel\\
$^{153}$ Raymond and Beverly Sackler School of Physics and Astronomy, Tel Aviv University, Tel Aviv, Israel\\
% ...
$^{158}$ Department of Physics, University of Toronto, Toronto ON, Canada\\
% ...
\end{flushleft}

\end{document}